\documentclass[a4paper]{mn2e}

\usepackage{epsfig}
\usepackage{graphics}
\usepackage{color}
\newcommand{\gsim}{\mathrel{\raise0.35ex\hbox{$\scriptstyle >$}\kern-0.6em % Greater/squiggles
\lower0.40ex\hbox{{$\scriptstyle \sim$}}}}
\newcommand{\lsim}{\mathrel{\raise0.35ex\hbox{$\scriptstyle <$}\kern-0.6em % Less than/squiggles
\lower0.40ex\hbox{{$\scriptstyle \sim$}}}}

\def\ang{\mbox{\AA}}
\def\kms{\mbox{${\rm km~s}^{-1}$}}
\def\ergsec{\mbox{${\rm erg}\thinspace{\rm s}^{-1}$}}
\def\ergsqcmsec{\mbox{${\rm erg \thinspace cm}^{-2}{\rm s}^{-1}$}}

\def\asec{\hbox{$^{"}$}}
\def\msun{\mbox{${\rm M}_{\odot}$}}

\def\lsun{\mbox{${\rm L}_{\odot}$}}

\def\ale{\mathrel{\hbox{\rlap{\hbox{\lower4pt\hbox{$\sim$}}}\hbox{$<$}}}}
\def\age{\mathrel{\hbox{\rlap{\hbox{\lower4pt\hbox{$\sim$}}}\hbox{$>$}}}}
\def\etal{et~al.\ }

\begin{document}
\date{\today}

\title[A UCD in Sombrero]{An Ultra Compact Dwarf around the
Sombrero galaxy (M104): the Nearest Massive UCD}
\author[Hau \etal]{George~K.~T.~Hau$^{1}$, Lee R. Spitler$^{1}$, Duncan A. Forbes$^1$,
Robert N. Proctor$^1$, \newauthor Jay Strader$^2$,  J. Trevor Mendel$^1$,  Jean P. Brodie$^3$, William E. Harris$^4$
 \\
$^1$Centre for Astrophysics and Supercomputing, Swinburne University of Technology,
Hawthorn, Victoria 3122, Australia. \\
$^2$ Harvard-SmithsonianCfA, 60 GardenSt., Cambridge, MA 02144, USA. \\
$^3$ UCO/Lick Observatory, University of California, Santa Cruz, CA 95064, USA.\\
$^4$ Department of Physics and Astronomy, McMaster University, Hamilton, ON L8S 4M1, Canada  
 }
\maketitle

\begin{abstract}

We report the discovery of an Ultra Compact Dwarf (UCD)
associated with the Sombrero galaxy (M104).  This is the closest
massive UCD known and the first spectroscopically verified massive UCD which is located in a low density environment. 

The object, we name SUCD1, was identified in 
HST/ACS imaging and confirmed to be associated with the Sombrero
galaxy by its recession velocity obtained from Keck spectra. 
The light profile is well fitted by a Wilson model.
We measure a half light size of $14.7 \pm 1.4$ pc, an absolute magnitude
of $M_{\rm V}$ = $-12.3$ mag ($M_{\rm K}$ = $-15.1$ mag) and an internal velocity
dispersion of $25.0 \pm 5.6$  \kms. Such values are typical of UCDs.  
From Lick spectral indices we measure a luminosity-weighted central age of $12.6 \pm 0.9$ Gyrs, $[Fe/H]$ of $-0.08 \pm 0.08$ dex and  
$[\alpha/Fe]$ of $0.06 \pm 0.07$ dex. 
The lack of colour gradients suggests these values are representative of the entire UCD.
The derived stellar and
virial masses are the same, within errors, at 
$\sim$ 3.3 $\times$
10$^7$ M$_{\odot}$. Thus we find no strong evidence for dark matter or
the need to invoke a non-standard IMF. 

We also report arguably the first X-ray detection of a {\it bona fide} UCD, which we 
attribute to the presence of Low-Mass X-ray Binaries
(LMXBs). The X-ray luminosity of $L_X= 0.56 \times10^{38}$
\ergsec\ is consistent with the values observed for GCs of
the same metallicity.  Overall we find SUCD1 has properties similar to 
other known UCDs and massive GCs. 

\end{abstract}


\section{Introduction}

Ultra Compact Dwarfs (UCDs) are compact stellar systems that are
more luminous than typical globular clusters (GCs). 
They were discovered in the cores of galaxy clusters (Hilker
\etal 1999; Drinkwater \etal 2000).
They have properties intermediate between globular clusters (GCs) and
dwarf ellipticals (dEs) and may represent a
transitional population between the two (Ha{\c s}egan et al
2005). The four main hypotheses for UCD formation are: 1) 
extra luminous GCs (Mieske \etal 2002)
2) the products of the merger of super star clusters 
(Fellhauer \& Kroupa 2002); 3) the stripped nuclei of dwarf
ellipticals (Bekki \etal 2003) and 4) primordial dwarf galaxies 
(Drinkwater \etal 2004). The recent
work of Forbes et al. (2008) and references therein indicate that UCDs share many
of the properties of massive star clusters. 

As the UCDs identified so far reside mainly in clusters, 
the cluster environment
is thought to be important in UCD formation. 
Clearly the discovery of UCDs in
isolated environments or associated with individual galaxies will
provide challenges to any hypothesis that requires a cluster
environment. Evstigneeva \etal (2007a) found 1
definite and 4 possible UCD candidates in a photometric search in
6 galaxy groups. All of these are intergalactic and not
associated with any particular galaxy. 
Mieske, West \& Mendes de Oliveira (2007)
searched for UCDs photometrically in the NGC 1023 group
($D$ = 11 Mpc) 
yielding
21 candidates, which have not been spectroscopically confirmed.
Perhaps the most promising candidate so far has been the brightest globular cluster HCH99-18 (Rejkuba \etal 2007) of NGC 5128 (Cen A; $D$ =  3.8 Mpc), which has a mass of $1.1\times 10^7$ \msun\ (Mieske \etal 2008).
Here we present HST/ACS and Keck/DEIMOS and Keck/LRIS
observations to establish the association between a massive UCD with the
Sombrero galaxy, which is located in a low density environment.
We adopt 
a Sombrero distance of 9.0 Mpc (Spitler \etal 2006);
parameters from the literature have been automatically converted
to this system.

\section{UCD identification and kinematic parameters}
\label{sec:disp}

\begin{figure}
\includegraphics[scale=0.38]{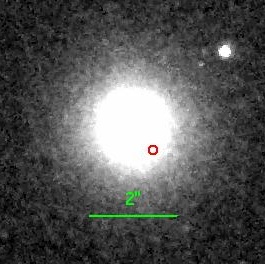}
\includegraphics[scale=0.28]{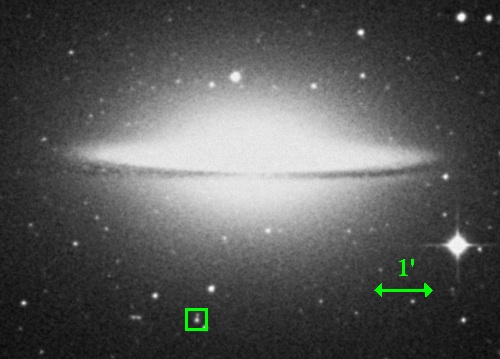}
\caption{Left: V-band HST/ACS image of the UCD. An error circle
of 0.1\asec radius (di Stefano \etal 2003) for the X-ray position
is marked by the red circle.  The scale bar is 2\asec. 
Right: Digital Sky Survey image showing the location of SUCD1 (square) with
respect to the Sombrero. The scale bar is 1\arcmin. North is up and East is left on
both images.  
Note the box on the right image is larger
than the area covered by the left image.  }
\label{figure:ucd}
\end{figure}

A mosaic of HST/ACS images of the Sombrero galaxy in the F435W (B), F555W
(V) and F625W (R) bands were taken as
part of the Hubble Heritage program. 
Spitler et al. (2006) have presented photometry of the Sombrero
GC system based on this data.
We identified a candidate
UCD based on its angular size (FWHM $\sim$ 0.15\asec), apparent
magnitude ($V=17.46$)
and colour ($B-V = 0.91$). It is $1.2$ magnitudes brighter
than the brightest Sombrero GCs in the V band (Spitler et al. 2006).  
An image of the UCD, as well as its location 3\arcmin\ (7.9
kpc) South of the
Sombrero, is shown in Fig.~\ref{figure:ucd}. The observed properties of the
UCD, which we name SUCD1, are summarised in Table ~\ref{tab:parameters}. 

\begin{figure}
\includegraphics[scale=0.32]{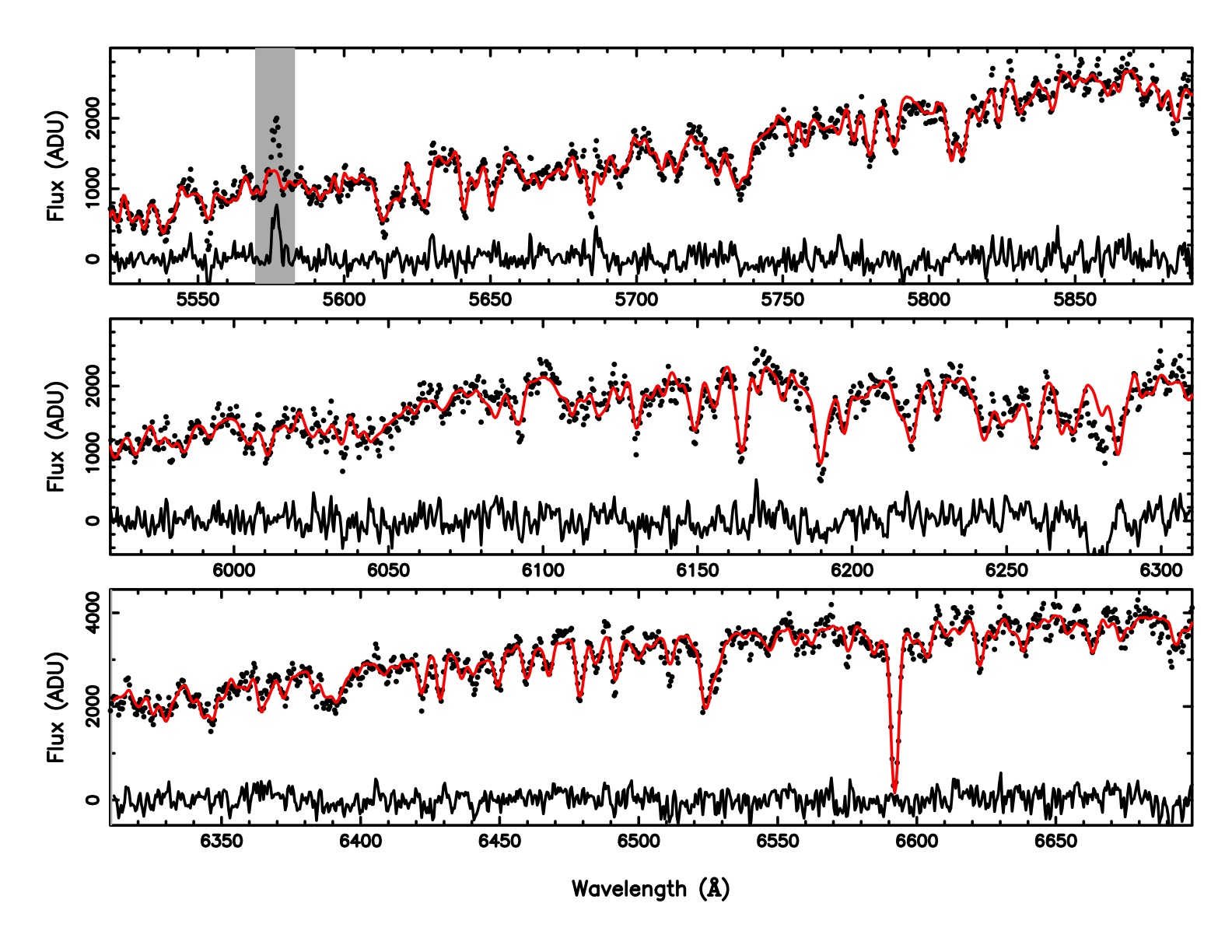}
\caption{DEIMOS spectrum of SUCD1 (black dots), and best-fit model (red lines). Each panel
corresponds to a different wavelength region. The residuals are plotted at the bottom of each panel. The grey area indicates masked regions.
For display purpose, the spectra have been shifted down by 4500,
6000 and 6100 ADUs (top to bottom). 
} 
\label{figure:deimos}
\end{figure}

A spectrum of SUCD1 was taken in 2006 April using the DEIMOS
instrument on the Keck telescope, with
12037 seconds exposure, GG400 order blocking
filter, 0.8\asec\ slit width and the 900 lines/mm grating. 
The spectral
range is 5500--7250 \AA\ and the resolution is $\sim0.9\ \ang$ or $51\ \kms$ in sigma.  
The
spectrum is plotted in Fig.~\ref{figure:deimos}. 
A lower resolution spectrum  ($\sim 1.2\ \ang$ or 74 \kms in sigma)  was taken using with the blue arm of
the LRIS instrument in 2008 April.
A total of
1800 seconds 
exposure was taken, with 1.0\asec\ slit width and
the 600/4000 grism.  The spectral range is 3600--5600\ \AA.

The recession velocity and velocity dispersion were measured from the
DEIMOS spectrum using a direct fitting program {\tt pixfitgau},
described in van der Marel \& Franx (1993).
The
region containing the NaI line at 5890 \AA\ was excluded from the
fits since it is affected by interstellar absorption. Four
template star spectra taken with DEIMOS were available for the
modelling. 
We find that no single star gives the best fit for the entire spectrum, and that the best-fit template changes with wavelength.
We therefore performed the spectral fitting separately on 3 separate
ranges corresponding to 5520--5890 \AA,
5960--6310 \AA, 6310--6700 \AA. For each
region, the best fitting parameters are taken for the broadened
template which gives the least $\chi^2$ between the model and the
data. We find that this method works well and the fits are
displayed in Fig.~\ref{figure:deimos}. The final parameters are
taken as the average obtained from the 3 regions, with the standard deviation taken as the error.

The measured heliocentric recession velocity $v_{helio}$ of $1293.1 \pm 9.5$
\kms\ confirms SUCD1's association with Sombrero, since the
recession velocity of the latter is $1024 \pm 5$ \kms (Smith
\etal 2000). 
 Bridges \etal (2007) 
found 
a dispersion of $233 \pm 20$ \kms\ in the GC population within
5\arcmin\ radius.  Given that SUCD1 is $\sim 3\arcmin$ from the
Sombrero, its relative velocity of $269 \pm 11\ \kms$ is
slightly larger than one standard deviation of the GC velocity
distribution.

The measured internal velocity dispersion $\sigma$ is $25.0 \pm 5.6\
\kms$, similar to other UCDs 
(Drinkwater et al 2003;
Ha\c{s}egan \etal 2005; Evstigneeva \etal 2007b; Hilker et al
2007). 
This is backed up by our Monte Carlo simulations which show that velocity dispersions as low as 4 \kms\ can be measured without significant bias (see also Bedregal 2006). 
We also measured the recession velocity from the
LRIS spectrum with a cross-correlation program {\tt fxcor} (Tonry
\& Davis 1979), and obtained $1305.0 \pm 9.1$ \kms. The agreement between instruments and techniques is reassuring.

\begin{table}
\begin{center}
\begin{tabular}{lr}
\hline 
Sombrero UCD (SUCD1) & Value \\
\hline
$\alpha^a$ (J2000) &  12 40 03.13 \\ %checked
$\delta^a$ (J2000) & -11 40 04.3 \\ 
Distance$^{b}$ & $9.0$ Mpc\\  % checked
Projected scale$^{b}$ & 44 pc arcsec$^{-1}$\\ % this is for d=9 Mpc
$V$$^{g}$  & $17.46$ mag \\
$B-V$$^{g}$ & $0.91$ mag \\
$B-R$$^{g}$ & $1.50$ mag\\
$V-R$$^{g}$ & $0.58$ mag\\
$J$ $^{e}$ & 15.6 mag \\ 
$H$$^{e}$ &  14.9 mag\\ % checked
$K$$^{e}$ & 14.7 mag \\ % checked
$3.6\ \mu$m$^{c}$ & 14.2 mag \\
$M_V$ & $-12.31$ mag \\ % using V=17.46 & m-M = 29.77
$L_V$ & $6.98 \times 10^6$ \lsun \\ % using M_Vsun=4.8
$M_K$ & $-15.1$ mag \\ % using m-M=29.77 and K=14.7
$L_X$$^f$ & $0.56 \times 10^{38}$ \ergsqcmsec \\
$A_B$$^{d}$ & 0.22 mag \\  % checked
\hline
\end{tabular}
\end{center}
Notes: 
$^a$~from ACS image calibrated using 2MASS catalogue; 
$^b$~Spitler \etal (2006);
$^c$~Spitler \etal (2008);
$^d$~Schlegel \etal (1998);
$^e$~2MASS;
$^f$~Di Stefano \etal (2003);
$^g$~Extinction corrected.
\caption{Table of properties of SUCD1. 
}
\label{tab:parameters}
\end{table}

\section{Stellar populations}\label{sec:HII}

The wavelength range of 4300--6300 \ang\ covered by the LRIS spectrum includes 16 Lick/IDS indices (Worthey et al. 1994). The indices were used to derive luminosity-weighted log ($age$), iron abundance $[Fe/H]$ and $\alpha$-abundance ratio $[\alpha/Fe]$ using the method described in Proctor \& Sansom (2002).  The extraction aperture is $44\times72$ pc.
Thomas et al. (2003) SSP models and Trager \etal (1998) index
definitions have been adopted. The measured log ($age$), $[Fe/H]$
and $[\alpha/Fe]$ are $1.10 \pm 0.03$ dex (12.6 Gyrs), $-0.082\pm
0.081$ dex and $0.06\pm0.07$ dex respectively. These are consistent with the values gleaned from photometric colour-colour analysis (Spitler \etal 2008).
The $[Fe/H]$ is comparable to those of very metal-rich GCs (Beasley \etal 2008;
Proctor \etal 2008; Norris
\etal 2008).
It is generally consistent with UCDs of similar mass or luminosity (e.g. Evstigneeva \etal 2007b; Chilingarian \& Mamon 2008; Dabringhausen, Hilker \& Kroupa 2008, Mieske \etal 2008). 
 It is higher than
dE nuclei which have $[Fe/H] \sim -0.3 \pm 0.1$ dex in the Virgo cluster (Geha \etal 2003), or $\sim -1.4 \pm 0.2$ dex in the Fornax cluster (Mieske \etal 2006). SUCD1's age is comparable to those of very metal-rich GCs and some Virgo UCDs, but is higher than the average age of 5 $\pm$ 3 Gyr for Virgo dE nuclei.

\section{Photometric properties}

\begin{figure}
\includegraphics[scale=0.45]{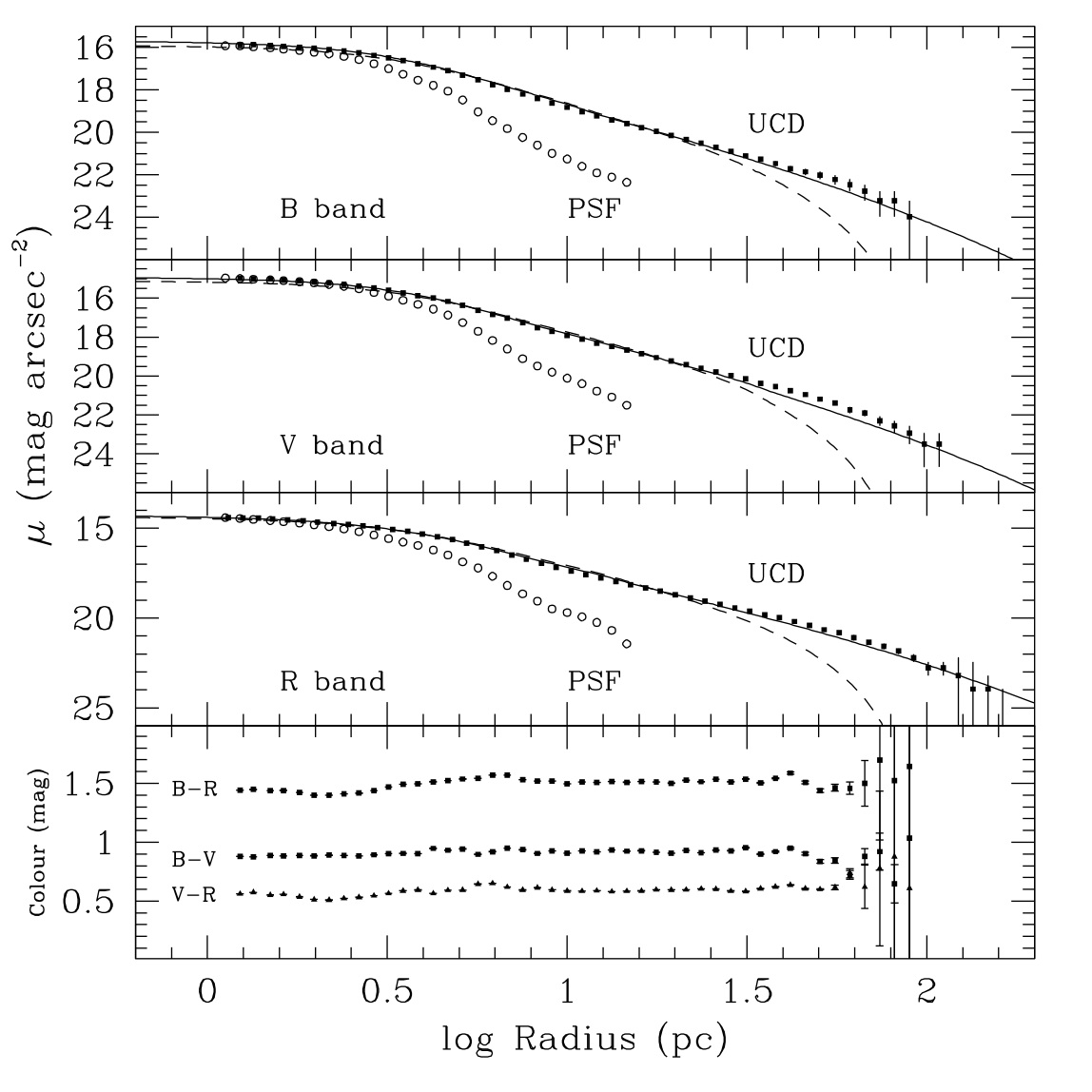}
\caption{Top 3 panels: Surface brightness profiles of SUCD1 in B, V and R bands, measured directly from the image (solid points). The fitted  Wilson and King models, convolved with the PSF, are represented by the solid and dashed curves respectively. The PSF is represented by the open circles. 
Bottom: Colour profiles of SUCD1: $B-R$,  $B-V$ and $V-R$.  No strong colour gradient is observed.
}
\label{figure:sb}
\end{figure}

We have also carried out profile fitting using the
equilibrium dynamical models of King (1966)
and Wilson (1975) in the form described by
McLaughlin \& van der Marel (2005), convolved with the PSF
in each case.  {\tt STSDAS/ELLIPSE} was used to measure
the surface brightness of SUCD1 out to large 
radii, and the two nearest moderately bright stars were used to define the PSF in each of B, V, R bands.  The observed profiles
are shown in Figure \ref{figure:sb}.  In all three cases SUCD1 was found to be quite round ($e \ale 0.05$) and
so the assumption of a one-dimensional circularly
symmetric profile for the convolved fits is reasonable.
The technique is described in McLaughlin \etal (2008).

The best-fitting Wilson-type models are in every case far superior to the King-type
models.  The reasons for this are well described
in McLaughlin \& van der Marel (2005) and McLaughlin \etal (2008),
but are due essentially to the different assumption the Wilson
model makes to treat the stars near the escape energy, leading
to a larger envelope and larger formal tidal radius.  
The solutions from the three filters are entirely self-consistent,
giving Wilson parameters of central potential $W_0 = 8.9$, concentration $c = 3.329$,
and scale radius $r_0 = 0.57$ pc.  The equivalent
projected half-light radius $r_h$ (of the \emph{intrinsic} profile
after deconvolution) found from direct integration of the
intrinsic surface brightness curve is $14.7 \pm 1.4$ pc.
We do not find any substructure or tidal tails in the
residual image.

SUCD1's colour profiles are also plotted in
Fig.~\ref{figure:sb}. The stellar population analysis above predicts 1.61, 1.01 and 0.69 for  $B-R$, $B-V$ and $V-R$ respectively. The measured colours in this work agree with the predictions to within 0.1 mag.
The lack of colour gradients is 
largely consistent with the findings for other UCDs (Evstigneeva
\etal 2008), and suggests that the
stellar population is largely uniform out to $\sim$ 50 pc, i.e. SUCD1 is globally old with slightly subsolar metallicities and near solar $\alpha$ element ratio. 

\section{Mass and mass-to-light ratio}
\label{sec:mass}

\begin{figure}
\includegraphics[scale=0.55]{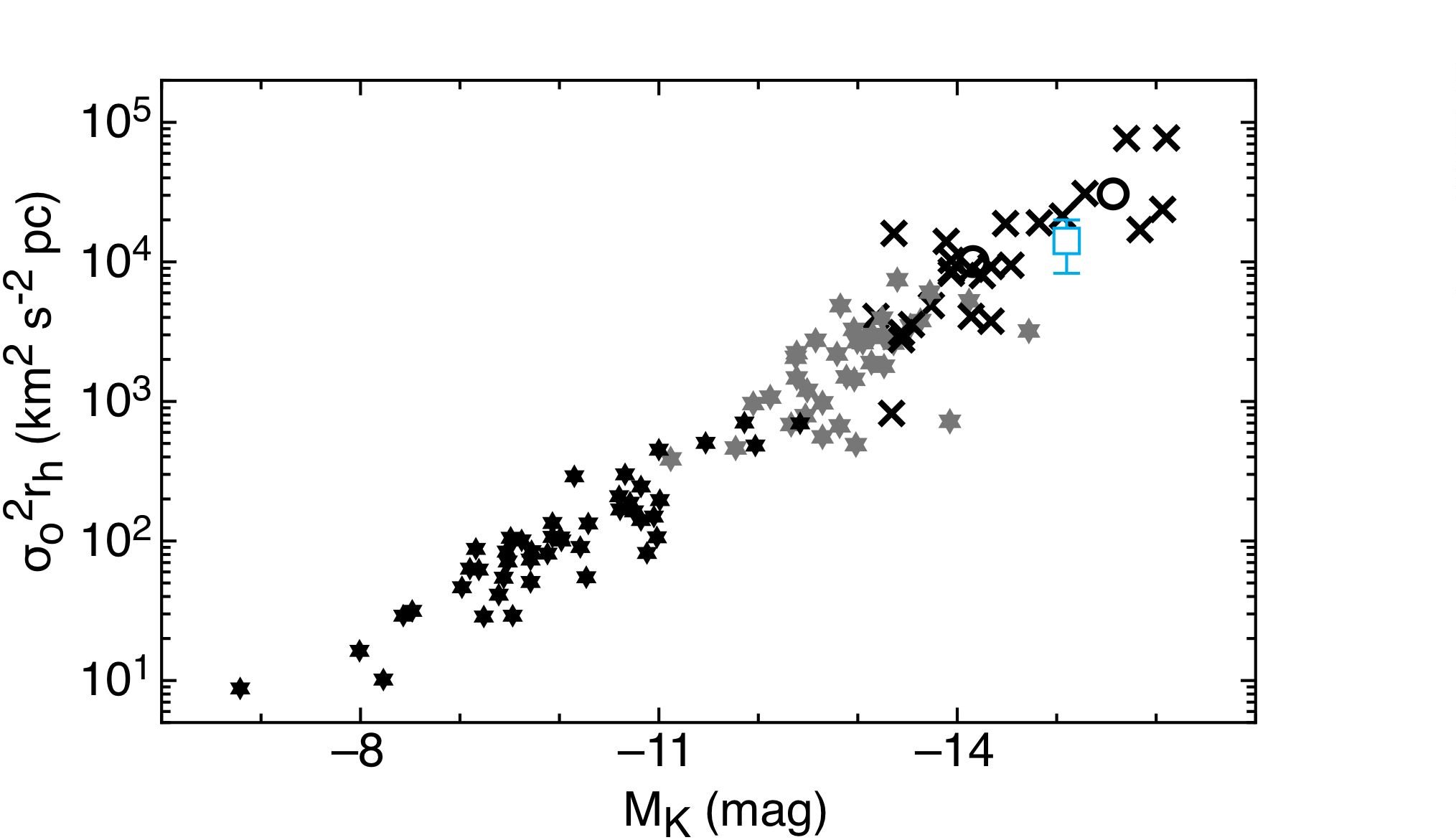}
\caption{Dynamical mass indicator $\sigma_0^2 R_h$ vs stellar mass indicator $M_K$ for GCs, UCDs and dE,N nuclei,
from Forbes \etal (2008). SUCD1 is indicated by the
large open square; Galactic and M31 GCs are shown by dark stars. NGC 5128 GCs by light stars. UCDs  and dE,N nuclei are marked by crosses and open circles respectively. SUCD1 follows the trend for other UCDs and massive GCs.}
\label{fig:sigma2r}
\end{figure}

To convert the measured $\sigma$ to $\sigma_0$, the central value, we use the conversion factor $\sigma_0/\sigma = 1.24$ calculated by Hilker \etal (2007) who modelled 5 Fornax UCDs. 
What is important in this conversion is ratio of the slit width to $r_h$. By coincidence this quantity is the same for SUCD1 as for the UCDs in the Hilker \etal sample. Thus we calculate $\sigma_0= 31.0 \pm 6.9$ \kms. In Fig.~\ref{fig:sigma2r} we show the location of SUCD1 with respect to
other UCDs, dE,N nuclei, GCs and massive star clusters in terms
of their $\sigma_0^2 r_h$ (a measure of dynamical mass) versus their K-band magnitude (a good 
proxy for stellar mass). SUCD1 falls in the location of other UCDs and the continuation of the massive GC trend.

Assuming a 
Kroupa IMF, the stellar mass calculated from the
K-band luminosity (using $M/L_K = 1.65$ from the stellar population analysis), 
is $3.6 \times 10^7$
\msun, while that from the  V-band (using $M/L_V =4.36$) is
$3.0\times10^7$ \msun. The virial mass $M_{virial} =
C \times \sigma_0^2 r_{h} / G$, where G is the universal
gravitational constant, and $C=10$ is the virial coefficient (for
more details see Forbes \etal 2008). Using the 
$\sigma_0$ calculated above,
and $r_h =14.7 \pm 1.4$~pc, we calculate 
$M_{virial} = 3.3 \pm 1.7 \times 10^7 \msun$. 
This  agrees
with those derived from the stellar populations. 
The resulting dynamical $M/L_{V} = 4.7 \pm 2.4$ is similar to those of
UCDs (Mieske \etal 2008) and
massive GCs (McLaughlin \etal 2008), and follows the trend of increasing $M/L$ with mass.
These results suggest that SUCD1 is not heavily dark matter
dominated (e.g. Dabringhausen \etal 2008; Forbes \etal 2008), nor does it require an exotic IMF (Mieske \& Kroupa 2008; Murray 2008), and it has not been strongly influenced by the host galaxy (Fellhauer \& Kroupa 2006).

\section{X-ray detection}  

SUCD1 is detected in X-rays with Chandra by di Stefano \etal (2003) and Li, Wang \& Hameed (2007), with IDs X95 and XA-143 respectively.
di Stefano \etal  identified X-ray
counterparts using a list of Sombrero GCs known
at that time, and listed SUCD1 as a star with $L_X$ (0.3-0.7 keV) of
$0.56\times 10^{38}$ \ergsec. 
The X-ray
position is 0.7\asec (31 pc) from the centre of SUCD1. The two
nearest stars are 2.7\asec and 5.4\asec from the X-ray position
and can be ruled out as the source of the X-ray emission at the
9$\sigma$ level.

SUCD1 is arguably the first {\it bona fide} UCD detected in the X-ray. 
Hempel \etal (2007) found an optically bright  X-ray source in the proximity of the field galaxy NGC 3585 (ID 18), with $L_X \sim 0.84 \times 10^{38} \ergsec$ and $B-K$ = 3.79 mag. 
Lacking spectroscopic confirmation,
if indeed it is at the distance of NGC 3585, this object will have $M_{\rm K} = -14.8$~mag, similar to SUCD1. Mieske \etal (2008) identified X-ray counterparts of Fornax compact stellar systems up to $M_{\rm V} \age -11.1$ mag, the arbitrary `boundary' between GCs and UCDs (Ha{\c s}egan \etal 2005; Mieske \etal 2006). 
With $M_{\rm V} = -12.3$ mag ($M_{\rm K} = -15.1$~mag), SUCD1 is firmly in the UCD regime. Its  $L_X$ is similar to those of  Low-Mass X-ray Binaries (LMXBs) in GCs (Kundu, Maccarone \& Zepf 2007; Mieske \etal 2008),
and about 1.5 dex lower than that of the most X-ray luminous GCs. 
This is  consistent with the finding that, in general, X-ray luminous GCs tend to be massive, but the brightest X-ray source is not necessarily the most massive GC. 
The preference of LMXBs to be located in metal-rich GCs (Jord\'an \etal 2004; Kundu, Maccarone \& Zepf 2007; Mieske \etal 2008; Woodley \etal 2008) is thought to be due to an increased number of neutron stars per unit mass. 

\section{Discussion and conclusions}

The observations reported here establish SUCD1 as a {\it
bona fide} UCD ($M_{\rm V}$ = $-12.31$, $r_h$ = 14.7 pc) 
associated with the Sombrero galaxy. Sombrero is in a low-density environment, and is listed as either an isolated galaxy or the dominant member of a small group in the literature, depending on the group-finding algorithm adopted.
The close association of SUCD1 with a relatively isolated galaxy suggests that some UCDs are formed in low-density environments. 

However our findings do not favor a recently stripped dwarf origin. 
Although simulations of a dwarf galaxy
being tidally stripped as it orbits the Milky Way are able to produce
objects which look like $\omega$ Cen (Bekki \& Freeman 2003), the failure to find any tidal extension or tails rules
out that SUCD1 is the nucleus of a {\it recently} stripped dwarf. If SUCD1 is the nucleus of a stripped dwarf, the stripping must have happened very early on. SUCD1's stellar populations are also somewhat incompatible with those of currently studied dE nuclei which tend to be younger on average.

Our findings favor a massive metal-rich GC scenario.  SUCD1's velocity
and location with respect to Sombrero are consistent with it being
part of its GC system. The $M/L$, old age and near-solar
metallicity of SUCD1 are similar to very metal-rich
GCs. SUCD1's photometric properties are also consistent with the average properties of the Sombrero metal-rich GC subpopulation (Spitler \etal 2008). 
The first discovery of X-ray emission in a spectroscopically confirmed UCD and its agreement with
LMXBs in GCs also supports the GC association, although it is possible that LMXBs also exist in dE nuclei. 
Our $M/L$ finding suggests that SUCD1 is largely consistent with  a purely stellar population
and is not heavily dark mater dominated, in agreement with the finding for some UCDs
by other works (Hilker \etal 2007; Dabringhausen, Hilker \&
Kroupa 2008; Forbes \etal 2008).  
We note that SUCD1 is 1.2 mag brighter in the V band than the next brightest Sombrero GC, which may argue that SUCD1 is not part of the GC system. However it is unclear whether this is due to low number statistics or incompleteness due to the limited ACS FOV coverage.

High-resolution cosmological simulations predict that, if GCs were formed in the cores of supergiant molecular clouds at high-redshifts,
then the maximum GC mass should correlate with the mass of its
host galaxy (Kravtsov \& Gnedin 2005). To explore the possibility that SUCD1 is just a very massive GC formed during the formation of the Sombrero galaxy, we use equation 8 of
Kravtsov \& Gnedin (2005) to calculate the mass of the most massive star cluster to be  $2.6 \times 10^7$ \msun, which compares
favourably with SUCD1's mass. A Sombrero mass of $5.4
\times10^{11}$ \msun\  was adopted (Bridges \etal 2007). 
It is unclear, however, whether this model is applicable to SUCD1, since it predicts low metallicities. Clearly further simulation work is required. 

Recently, Bailin \& Harris (2008) consider the self enrichment of globular clusters at high-redshifts. They find that metal retainment becomes efficient at high masses ($\sim 10^7$ \msun), and predict that the red and blue GC sequences should  converge there.
With a 0.4 star formation efficiency, their model is able to generate a $[Z/H]$ of $-0.03$ dex for a  $\sim 3\times 10^7$~\msun\ GC, consistent with that of SUCD1 
and other UCDs (Dabringhausen \etal 2008) of that mass.

The similarity of SUCD1's properties to those of cluster UCDs raises the intriguing possibility that some cluster  UCDs could have been formed first in low-density environs together with the GC population, and were then  incorporated into the cluster later. 
It may be possible that UCDs form from multiple routes, as hinted by the different stellar populations of the Virgo and Fornax UCDs found by others, and individual circumstances need to be examined to study the relative importance of different channels. Clearly the discovery of more UCDs associated with isolated galaxies will help to shed light on their formation.

\section{Acknowledgements}
GKTH and DAF
thank the Australian Research Council for financial support. This work is supported by NSF grant AST 05-07729. We thank Warrick Couch, Alister Graham and Paul Lasky and  for useful comments.

\end{document}